\documentclass[12pt]{article}
\usepackage{graphicx}
\usepackage{amssymb,amsmath,amsfonts,palatino,amsthm}
\usepackage{amssymb}
\usepackage{epstopdf}
\newcommand{\beq}{\begin{equation}}
\newcommand{\eeq}{\end{equation}}

\newcommand{\f}{\begin{equation}}
\newcommand{\ff}{\end{equation}}

\begin{document}

%%%%%%%%%%%%%%%%%%%%%%%%%%%%%%%%%%%%%%%%%%%%%%%%
\title{Positive energy in quantum gravity \\}
\author{Lee Smolin\thanks{lsmolin@perimeterinstitute.ca} 
\\
\\
Perimeter Institute for Theoretical Physics,\\
31 Caroline Street North, Waterloo, Ontario N2J 2Y5, Canada}
\date{\today}
\maketitle
%\vfill

\begin{abstract}
This paper addresses the question of whether Witten's proof of positive $ADM$ energy for classical general relativity\cite{Witten} can be extended to give a  proof of positive energy for a non-perturbative quantization of general relativity.
To address this question, a set of conditions is shown to be sufficient for showing the positivity of a Hamiltonian operator corresponding to the 
$ADM$ energy.  One of these conditions is a particular factor ordering for the constraints of general relativity, in a representation where the states are functionals of the Ashtekar connection, and the auxiliary, Witten spinor.  

%Witten's proof of positive energy is particularly simple in Ashtekar variables.  This allows us to to  show how Witten's spinor equation arises from the constraint that generates chiral supersymmetry transformations in supergravity.

These developments are partly based on results derived with Artem Starodubtsev\cite{ArtemLeepositive}.

\end{abstract}

\tableofcontents

\newpage

\section{Introduction}

One of the most evident facts about the world is the stability of empty space-time.  In classical general relativity we can explain this as a consequence of the positive energy theorem, which establishes, in the asymptotically flat context, that, when the constraints of the theory are satisfied, and matter satisfies the positive energy condition, the $ADM$ mass is positive definite.  Further, the $ADM$ mass only vanishes when the space-time is Minkowski space-time.  This theorem was proved first by Shoen and Yau\cite{SY}, although here we will be interested in a slightly later proof of Witten\cite{Witten}.

In this paper we discuss a corresponding result for the quantum theory of gravity.  Certainly the positive energy theorem must extend in some suitable form to any viable quantum theory of gravity.  This is highly non-trivial in a background independent approach  because, as a consequence of the equivalence principle, the $ADM$ hamiltonian comprises a bulk term, which is proportional to constraints, and a boundary term, which is not positive definite off the constraint surface.  

To make progress towards such a quantum positive energy theorem, we study a particular class of theories, where the quantum state is a functional of the Ashtekar connection (which is the chiral $SU(2)_{Left}$ part of the space-time connection) and an auxiliary spinor variable, the Witten spinor.   Working within this class of representations, we establish a set of sufficient conditions for a quantization of general relativity to have such a theorem.   To do this, we work at a formal level in which we pay attention to operator ordering, but not the details of a regularization scheme for operator products.

One crucial issue that emerges is the requirement that the spatial metric and frame fields be non-degenerate.  This is a necessary condition of the classical proof\cite{Madhavan}, and the quantum proof requires correspondingly that $\hat{\frac{1}{\sqrt{det (q)}}}$ be well defined as a quantum operator.  This is a challenge for the standard Ashtekar-Lewandowski representation of loop quantum gravity, which allows for states where the metric operators are degenerate.  This is indeed a crucial issue because the fact that the configuration space of the theory extends to degenerate three metrics is a consequence of the fact that the action, equations of motion and constraints of the connection based form of general relativity are all low order polynomials; the same circumstance which makes possible exact and non-trivial results in the quantum theory.  

As a byproduct of our work, we show that some known classical results have particularly simple derivations within the Ashtekar formalsm.  These include Witten's positive energy theorem itself and the demonstration that there exists a positive definite bulk Hamiltonian which is, however, only equal to the $ADM$ Hamiltonian on the constraint surface.

%We consider a class of theories based on the Ashtekar variables, but it should be pointed out at the outset that the standard representation mostly studied in loop quantum gravity (LQG)-that of Ashtekar and Lewandowski, does not satisfy two of the conditions we find.  The most crucial of these is that the quantum operators representing the components of the spatial metric, are invertible-not as operators on Hilbert space-but in the sense that they allow operators corresponding to the inverse of the determinant of the metric, to be defined.  We discuss in section () one way in which this may be remedied.

\subsection{Heuristic motivation}

The positive energy theorem was for half a century or more an open challenge to relativists.  Many attempts were made to prove flat spacerime was stable, but none completely succeeded completely until a majestic tour de force of geometric reasoning of Shoen and Yau\cite{SY}.  This was followed two years later by a proof of Witten\cite{Witten}, which was as elegant as it was short.  It is this proof of Witten's that we take as a template here for the quantum theory.  

Witten's proof was inspired by an observation about supergravity made by Grisaru\cite{Grisaru} and 
Deser and Teitelboim\cite{DT}.  This was that
the Hamiltonian of supergravity is positive definite because the $ADM$ Hamiltonian is the square of the supersymmetry charge.  In informal notation\footnote{In this paper indices $A,B,C, \dots = 0,1$ are left handed Weyl spinor indices,
while primed indices, $A',B',C', \dots = 0',1'$ signify the complex conjugate representation spanned by right handed Weyl spinors.   $a,b,c=1,2,3$ are three dimensional space-time indices.},
\f
H_{ADM} = {\cal Q}^{\dagger}_A {\cal Q}^A \geq 0
\label{Q2}
\ff

The suggestion was that a positive energy proof for general relativity could be gotten by restricting supergravity 
to its bosonic sector, which is general relativity.  Witten realized this suggestion in a very clever way which can be explained as follows.

Let us work in the chiral Hamiltonian formulation of $N=1$ supergravity, as presented by Jacobson\cite{Ted-sgr}.  There, both the Hamiltonian and the supersymmetry charge are a sum of a bulk term proportional to constraints and a surface integral take at the boundary at spatial infinity.  We will not need the full formulation here, but to motivate Witten's proof we need to know two things about it.  First, it extends the Ashtekar formulation of general relativity. Its canonical co-ordinates are the left handed part of the gravitational connection, or Ashtekar connection, $A_a^{AB}$, and the left handed gravitino field, $\psi^A_a$.  Their conjugate momenta are the densitized 
frame field, $\tilde{E}^a_{AB}$ and the gravitino momenta $\tilde{\pi}^a_A$.    The non-vanishing Poisson brackets are
\begin{eqnarray}
\{ A_a^{AB} (x) , \tilde{E}^a_{CD} (y) \} &=& \delta^3 (x,y) \delta_a^b \delta^{AB}_{CD}
\label{[A,E]}
\\
\{ \psi_a^{A} (x) , \tilde{\pi}^a_{C} (y) \}_+ &=& \delta^3 (x,y) \delta_a^b \delta_{A}^{C}
\label{[psi]}
\end{eqnarray}

Second, the constraint that generates left handed supersymmetry transformations has the form
\f
{\cal S}^A=  {\cal D}_a \pi^{aA }   =0
\ff
where $ {\cal D}_a$ is the left handed part of the gravitational connection, known as the Ashtekar connection.

The general relativity sector of supergravity can be taken to be the configurations in which the spinor field, 
$\psi^A_a$ and its conjugate momenta, $\pi^a_B$ vanish.  But there is a larger sector of the phase space which is gauge equivalent to general relativity under local supersymmetry transformations.  The left handed part of this is 
\f
\psi^A_a \rightarrow \delta_\xi \psi^A_a = \{  \psi^A_a ,  {\cal S}(\xi ) \} = {\cal D}_a \xi^A
\ff
To fully parametrize this sector of supergravity, which is gauge equivalent to general relativity, in a way that gets as close to preserving the Poisson brackets as possible, we may try to take,
\f
\tilde{\pi}^a_A \rightarrow \tilde{E}^a_{AB} \xi^{B}
\label{pireduce}
\ff
Then\footnote{If we want to preserve the precise Poisson bracket we should take, instead of (\ref{pireduce})
\f
\tilde{\pi}^a_A \rightarrow  \frac{1}{\xi_{E}\xi^{E}} \tilde{E}^a_{AB} \xi^{B}
\nonumber
\ff
but this runs afoul of the fact that Grassmann numbers don't have inverses.  To make sense of this we could try to extend the Grassmann algebra to a non-associative algebra, but this is too much novelty for a peripheral point.
}
\f
\{ \psi_a^{A} (x) , \tilde{\pi}^a_{C} (y) \}_+ \rightarrow \xi_E \xi^E  \delta^3 (x,y) \delta_a^b \delta_{A}^{C}
\ff

Then the supersymmetry constraint ${\cal S}^E$ becomes an elliptic equation for $\xi^E$
\f
{\cal S}^E \rightarrow {\cal G}^{EF}\xi_F + {\cal W}(\xi)^E =0 
\ff
where here 
\f
{\cal W}_E = \tilde{E}^a_{EF} {\cal D}_a \xi^F =0 
\label{Witteneq}
\ff
is known as the Witten equation, as it plays a key role in Witten's proof.

The other term in the equation is
\f
{\cal G}_{AB}^{gr} = {\cal D}_a \tilde{E}^a_{AB} =0  
\label{Gauss}
\ff
which is the Gauss law constraint that generates local chiral $SU(2)_L$ frame rotations.

To complete the description of this sector we may add a conjugate momenta $\pi_E $ to the theory, satisfying
\f
\{ \xi)_{A} (x) , \tilde{\pi}^{C} (y) \}_+ = \delta^C_A \delta^3 (x,y )
\ff
This doesn't play much of a role, except in one place below.

Let us call this sector of supergravity the {\it bosonic sector of supergravity.}  It is locally super-gauge equivalent to general relativity, although it might have novel topological effects.  

An appropriate restriction of the supercharge squared in (\ref{Q2}) to this sector gauge equivalent to general relativity is then  to square the Witten equation.  This is the starting point of Witten's proof, which is reproduced in the next section.

If we seek to extend the positive energy proof of Witten to the quantum theory, the first question to be confronted is what is the appropriate way to represent the Witten spinor and its equation in the Hilbert space?

A first thought (which was investigated in \cite{ArtemLeepositive}) is to take the spinor as an operator on the quantum gravity Hilbert space. This means to solve the Witten equation as a strong operator equation
\f
\hat{\cal W}_E = \tilde{E}^a_{EF} {\cal D}_a \hat{\xi}^F =0 
\ff
which when solved expresses $\hat{\xi}^F =\hat{\xi}^F (\hat{A}, \hat{E} )$ as a (very) non-linear and non-local functional of the 
gravitational operators $\hat{A}$ and  $\hat{E}$.  However, it turns out that because of operator ordering issues in the proof,
the spinor operator $\hat{\xi}^F$ would have to commute with the operators that represent the Hamiltonian and diffeomorphism constraints and so be what is called a Dirac observable.  Given that the Witten equation does not commute  with those constraints this seems to be too much to ask. 

So we try here something different, which is to put the Witten spinor into the wave functional, so that quantum states are functionals of $A_a^{AB}$ and $\xi^E$.
\f
\Phi = \Phi [ A_a^{AB} , \xi^E ]
\ff 
This  doesn't change the number of degrees of freedom because the wave functionals are subject to an additional pair of constraints-the Witten constraint,
\f
\hat{\cal W}^E  \Phi [ A_a^{AB} , \xi^E ] =0.
\label{Wittenc}
\ff
This can be thought of two ways.  First, we are used in theories with gauge invariance to writing quantum states on wavefunctionals on configuration spaces with auxiliary variables, which are then restricted to a dependence on the physical degrees of freedom by constraint equations.  This is just one more instance of it.  
%Indeed, as I show below, the Witten equation generates a set of gauge transformations of general relativity (or more properly- the bosonic sector of supergravity) that might be called its hidden supersymmetry.

We can also understand the quantum states of the form $ \Phi [ A_a^{AB} , \xi^E ]$ as a restriction to the bosonic sector of quantum supergravity.  

This however raises a difficult issue, which is that the first class nature of the constraint algebra is lost during the reduction from ${\cal S}^E$ to ${\cal W}^E$.  As just mentioned,  ${\cal W}^E$ fails to Poisson-commute with the usual constraints of general relativity.  This means that the others cannot be imposed as constraints on states as is usually done in loop quantum gravity.  Instead, the positive energy proof demands a weaker condition which is that the constraints-when 
smeared with a particular lapse and shift constructed from	 $\xi^E$, have vanishing expectation value.  

This brings us to the statement of the main result.  After this in section 2, I present Witten's classical proof of the positivity of the $ADM$ energy, expressed in Ashtekar variables\cite{ArtemLeepositive}.  In section 3, I a sketch of a translation of the classical proof into the quantum context.  

\subsection{Statement of the main result}

The main result of this paper is a set of sufficient conditions that a quantization	of general relativity must satisfy to have an operator representing the $ADM$ energy whose expectation values are positive.

Consider a representation of quantum general relativity whose states are functionals of the Ashtekar connection and the auxiliary spinor variables, $\xi^E$,
\f
\Phi = \Phi ( A^{AB}_a, \xi^E )
\ff
defined by the usual Ashtekar relations
\f
\hat{\tilde{E}}^a_{AB} \Phi [A,\xi ] = -\hbar \frac{\delta}{\delta A_a^{AB}}   \Phi [A,\xi ], \ \ \ \ \ 
\hat{A}^a_{AB} \Phi [A,\xi ] = A_a^{AB}  \Phi [A,\xi ]
\ff
together with operators for the spinor $\xi^E$ and its conjugate momenta $\tilde{\pi}_C$
\f
\hat{\tilde{\pi}}_{B} \Phi [A,\xi ] = -\hbar \frac{\delta}{\delta \xi^B}   \Phi [A,\xi ], \ \ \ \ \ 
\hat{\xi}^E \Phi [A,\xi ] = \xi^E  \Phi [A,\xi ]
\ff
which satisfies the following conditions
\begin{enumerate}

\item{}The inner product is defined by
\f
< \Phi (A, \xi ) | \Psi (A, \xi ) >= \int dA d\bar{A} d\xi d\bar{\xi} \ \bar{\Phi} (\bar{A}, \bar{\xi} ) e^{I(A,\bar{A},\xi, \bar{\xi})}  \Psi (A, \xi )
\ff
where $I(A,\bar{A},\xi, \bar{\xi})$ satisfies three conditions.  The first two are reality conditions for the frame fields and their time derivatives, while the third is a positivity condition for a certain operator.
\begin{eqnarray}
\frac{\delta e^{I(A,\bar{A},\xi, \bar{\xi})} }{\delta \bar{A}_{a}^{A'B'}(x) }  n^{A'A}  n^{B'B }
-   \frac{\delta e^{I(A,\bar{A},\xi, \bar{\xi})}}{\delta A_{a}^{AB}(x) }  &=& 0
\label{R1}
\\
n^{B'B} \nabla_a [   \hat{\frac{1}{e}} \frac{\delta}{\delta \bar{A}_{[a}^{A'B'}(x) }  \frac{\delta}{\delta A_{b]}^{AB}(x) } 
e^{I(A,\bar{A},\xi, \bar{\xi})}  ] &=&0
\label{R2}
\\
{\cal Q}^{ab}_{B'B} \equiv n^{A'A} \frac{\delta}{\delta \bar{A}_{(a}^{A'B'}(x) }  \frac{\delta}{\delta A_{b)}^{AB}(x) } 
e^{I(A,\bar{A},\xi, \bar{\xi})}  &>&0
\label{Q}
\end{eqnarray}
Here $n^{AA'}= n^a \sigma_a^{AA'}$ is a timeline unit normal such that $n^a n_b = n^a n^b \eta_{ab}=-1$.  

\item{}The quantum Witten equation holds as a constraint on states
\f
\hat{\cal W}^A  \Phi [A,\xi ] =  \frac{\delta}{\delta A_a^{AB}} {\cal D}_a \xi^B \Phi [A,\xi ] =0
\label{Wittenc2}
\ff   

We impose the boundary condition that as we go to infinity, the 
$\xi^E$ approaches a constant spinor $\lambda^E$ such that
\f
\bar{\lambda}^{E'} \lambda^E = s^{E'E}. 
\label{nullb0}
\ff
where $s^{A'A} = s^a \sigma^{AA'}_a$ is a constant future pointing null vector that is normalized to 
\f
s^a n_a =-1
\label{s^an_a}
\ff.

\item{} The $E^a_{AB}$ define an invertible metric, so that $\frac{1}{e}$ is a well defined operator.

\item{}The {\it expectation value of the scalar and vector quantum constraints hold}, when smeared against particular
lapse and shift constructed as follows from the Witten spinor
\f 
< \Phi | \int_\Sigma \bar{\xi}^{A'} n^{\ A}_{A'} \hat{\cal C}_{AB} \xi^B |\Phi > = 0 .  
\ff
\label{<C>}
in a particular ordering 
\f
\hat{\cal C}^{AB} = \hat{\tilde{E}}^{a A}_C  \hat{\tilde{E}}^{a C}_D  F_{ab D}^B .
\ff

The equivalence of these four constraints to the usual form of the Ashtekar  constraints, for non-degenerate three geometries,  was shown first by Jacobson in \cite{Ted-sgr}.

\end{enumerate}

The main result is then that when these conditions are satisfied the expectation value of the $ADM$ Hamiltonian for the null translation at infinity generated by $s^{A'A}$, is positive definite, where 
\begin{eqnarray}
<M_{ADM} > &=&- \int dA d\bar{A} d\xi d\bar{\xi}   \int_{\partial \Sigma} d^2 \sigma_a  ( n^{D}_{ B'} e^{I(A,\bar{A},\xi, \bar{\xi})}) \ 
(    \bar{\xi}^{B'}  \bar{\Phi} [ \bar{A}, \bar{\xi} ] ) \frac{1}{e} 
(  \frac{\delta}{\delta {A}_{[a A}^{D}(x) }  \frac{\delta}{\delta A_{b]}^{AB}(x) } {\cal D}_b \xi_B \Phi [A,\xi ] )
\nonumber
\\
%&=&- \int dA d\bar{A} d\xi d\bar{\xi}   \int_{\partial \Sigma} d^2 \sigma_a  \frac{1}{e} 
%( n^{D}_{ B'} e^{I(A,\bar{A},\xi, \bar{\xi})}) \ 
%(      \bar{\Phi} [ \bar{A}, \bar{\xi} ] )
%(  \frac{\delta}{\delta {A}_{[a A}^{D}(x) }  \frac{\delta}{\delta A_{b]}^{AB}(x) }A_{b B}^D  \Phi [A,\xi ] 
%\nonumber
%\\
& \geq &  0
\end{eqnarray}

\section{Classical proof of positive energy}

We first present Witten's proof of positive ADM energy, translated into chiral Ashtekar variables\footnote{This was 
done first in \cite{ArtemLeepositive}.}.

We start by squaring the Witten equation
\begin{eqnarray}
0=R&=& \int_\Sigma   
 \frac{n^{A'A}}{e} \bar{\cal W}_{A'} {\cal W}_A
\nonumber
\\
&=& \int_\Sigma \frac{n^{A'A}}{e}  \bar{\tilde{E}}^a_{A'B'}\bar{\cal D}_a \bar{\xi}^{B'} \tilde{E}^b_{AB}{\cal D}_b \xi^{B}
\end{eqnarray}
Note that the $\frac{1}{e}$ is necessary because the Witten equation, (\ref{Witteneq}),  inherits a density weight of one from that of the
$\tilde{E}^a_{AB}$.

In the presence of the Gauss's law constraint, ${\cal G}^{AB}_{gr}$ this is equivalent to squaring the supersymmetry generator
\f
0=R \approx  \int_\Sigma \frac{n^{A'A}}{e} \bar{\cal S}_{A'} {\cal S}_A
\ff
%Note that ${\cal G}^{AB}_{extended} \xi_B ={\cal G}^{AB}_{gr} \xi_B $

We can divide $R$ into symmetric and antisymmetric parts.
\f
R=R^{sym}+ R^{anti} =0
\ff
where,
\f
R^{sym} = \int_\Sigma \frac{n^{A'A}}{e} \bar{\tilde{E}}^{(a}_{A'B'}\bar{\cal D}_a \bar{\xi}^{B'} \tilde{E}^{b)}_{AB}{\cal D}_b \xi^{B} = 
 \int_\Sigma n_{B'B}{e} q^{ab}  \bar{\cal D}_a \bar{\xi}^{B'} {\cal D}_b \xi^{B} 
 \geq 0
\ff
is positive definite.

We then turn our attention to the antisymmetric part
\f
R^{anti} = \int_\Sigma \frac{n^{A'A}}{e} \bar{\tilde{E}}^{[a}_{A'B'} \bar{\cal D}_a \bar{\xi}^{B'} \tilde{E}^{b]}_{AB}{\cal D}_b \xi^{B} \leq 0
\ff
We note the reality conditions
\f
 n^{A'}_A  \bar{\tilde{E}}^{a}_{A'B'} n^{B'}_B = \tilde{E}^{a}_{AB}
\ff
and
\f
n_{B'}^B \nabla_a [\tilde{E}^{[a\ B'}_{A'} \tilde{E}^{b] \ C}_B   ] =0 
\ff
We make an integration by parts
\f
R^{anti} = 
\int_{\partial \Sigma} d^2 \sigma_a \mu^a - 
\int_\Sigma \frac{n^{A'A}}{e}  \bar{\xi}_{A'} {\cal C}_A^C \xi_{C} \leq 0
\ff
where 
\f
\mu^a=  \frac{n^{A'A}}{e}  \bar{\xi}_{A'}  [\tilde{E}^{[a} \tilde{E}^{b]} ]_{AB} {\cal D}_b \xi^B
\ff
and
\f
{\cal C}_A^C   = [\tilde{E}^{[a} \tilde{E}^{b]} ]_{AB} F_{ab}^{BC} =0
\ff
are four equations, equivalent to the four Ashtekar constraints.  When they are satisfied we have 
\f
- \int_{\partial \Sigma} d^2 \sigma_a \mu^a
= -  \int_{\partial \Sigma} d^2 \sigma_a \frac{n^{A'A}}{e}  \bar{\xi}_{A'}  [\tilde{E}^{[a} \tilde{E}^{b]} ]_{AB} {\cal D}_b \xi^B
 \equiv M_{ADM} \geq 0
 \label{ADM1}
\ff

Also, in the presence of the constraints, we have a positive definite expression for the null $ADM$ mass\footnote{Ted Jacobson has derived this expression directly\cite{Tedun}.}.
\f
M_{ADM} =R^{symm} =  \int_\Sigma n_{B'B}{e} q^{ab}  \bar{\cal D}_a \bar{\xi}^{B'} {\cal D}_b \xi^{B} 
 \geq 0
\ff
Three comments are in order.

\begin{enumerate}

\item{} The argument must be completed by a proof that the Witten equation (\ref{Witteneq}) has solutions asymptotic to any fixed null spinor at spatial infinity.  This is supplied by Witten\cite{Witten}, to which I have nothing to add.

\item{}To derive the positivity of the more usual timelike $ADM$ energy we need two spinors, $\xi^A_I$, where $I=1,2$,  each a solution to the Witten equation, chosen so that 
instead of (\ref{nullb0}),  we require that at infinity $\xi^A_I$ approach fixed spinors 
$\lambda^E_{I}$, such that
\f
\sum_I \bar{\lambda}^{E'}_I \lambda^E_I = n^{E'E}. 
\label{nullb}
\ff

\item{} If we now impose the standard fall off conditions on $\tilde{E}^a_{AB}$ and $A_a^{AB}$ then, as shown in \cite{Abhay-book},  (\ref{ADM1}) is equal to the standard $ADM$ mass.  However, it is important and interesting to note that even when less stringent boundary  conditions are imposed (\ref{ADM1}) still holds; only now what is proved to be positive is a highly non-linear expression, which we may call the generalized $ADM$ energy.  

\end{enumerate}

\section{Quantum positive energy}

Our aim in the following is to {\it find conditions a representation of quantum gravity may satisfy which are suffixient to guarantee the positive definiteness of an operator for the $ADM$ mass.} 

We begin again by squaring the Witten equation, only now we use the quantum version.
\begin{eqnarray}
0=<R>&=& \int_\Sigma d^3 x n^{A'A}< \bar{\cal W}_{A'}(x)  \bar{\Phi} (A, \xi ) |  \frac{1}{e} | {\cal W}_A (x)\Phi (A, \xi ) >
\\
&=& \int dA d\bar{A} d\xi d\bar{\xi}  e^{I(A,\bar{A},\xi, \bar{\xi})} \  \int_\Sigma d^3 x \frac{n^{A'A}}{e} 
( \frac{\delta}{\delta \bar{A}_a^{A'B'}(x) }  \bar{\cal D}_a   \bar{\xi}^{B'}  \bar{\Phi} [ \bar{A}, \bar{\xi} ] )
( \frac{\delta}{\delta A_b^{AB}(x) } {\cal D}_b \xi^B \Phi [A,\xi ] )
\nonumber
\end{eqnarray}

Again we divide into symmetric and antisymmetric parts
\f
<R>=<R^{sym}>+ <R^{anti}> =0
\ff

We want to show that the symmetric part is again positive definite.  To do this we integrate functionally by parts twice,
and use (\ref{R1}), to find,

\begin{eqnarray}
<R^{symm}> &=& \int dA d\bar{A} d\xi d\bar{\xi}  e^{I(A,\bar{A},\xi, \bar{\xi})} \  \int_\Sigma d^3 x \frac{n^{A'A}}{e} 
( \frac{\delta}{\delta \bar{A}_{(a}^{A'B'}(x) }  \bar{\cal D}_a   \bar{\xi}^{B'}  \bar{\Phi} [ \bar{A}, \bar{\xi} ] )
( \frac{\delta}{\delta A_{b)}^{AB}(x) } {\cal D}_b \xi^B \Phi [A,\xi ] )
\nonumber
\\
&=& \int dA d\bar{A} d\xi d\bar{\xi}  \int_\Sigma d^3 x \frac{n^{A'A}}{e}  ( \frac{\delta}{\delta \bar{A}_{(a}^{A'B'}(x) }  \frac{\delta}{\delta A_{b)}^{AB}(x) } e^{I(A,\bar{A},\xi, \bar{\xi})})  \ 
 ( \bar{\cal D}_a   \bar{\xi}^{B'}  \bar{\Phi} [ \bar{A}, \bar{\xi} ] )
({\cal D}_b \xi^B \Phi [A,\xi ] )
\nonumber
\\
&=& \int dA d\bar{A} d\xi d\bar{\xi}  \int_\Sigma d^3 x \frac{1}{e}  \  {\cal Q}^{ab}_{B'B} 
 ( \bar{\cal D}_a   \bar{\xi}^{B'}  \bar{\Phi} [ \bar{A}, \bar{\xi} ] )
({\cal D}_b \xi^B \Phi [A,\xi ] )   \geq 0
\label{<symm>}
\end{eqnarray}
where
\f
{\cal Q}^{ab}_{B'B} \equiv n^{A'A} \frac{\delta}{\delta \bar{A}_{(a}^{A'B'}(x) }  \frac{\delta}{\delta A_{b)}^{AB}(x) } 
e^{I(A,\bar{A},\xi, \bar{\xi})}
\ff
We now require that ${\cal Q}^{ab}_{B'B}   $ be a positive Hermitian matrix, which is (\ref{Q}).  
This implies that  (\ref{<symm>}) is positive definite.  

We then study the antisymmetric part:
\f
<R^{anti}> = \int dA d\bar{A} d\xi d\bar{\xi}  \int_\Sigma d^3 x \frac{n^{A'A}}{e} e^{I(A,\bar{A},\xi, \bar{\xi})} \   
(  \frac{\delta}{\delta \bar{A}_{[a}^{A'B'}(x) }  \bar{\cal D}_a   \bar{\xi}^{B'}  \bar{\Phi} [ \bar{A}, \bar{\xi} ] )
( \frac{\delta}{\delta A_{b]}^{AB}(x) } {\cal D}_b \xi^B \Phi [A,\xi ] ) \leq 0
\ff
We then functionally integrate by parts twice, but in  a different way
\begin{eqnarray}
<R^{anti}> &=& - \int dA d\bar{A} d\xi d\bar{\xi}   \int_\Sigma d^3 x \frac{n^{A'A}}{e} ( \frac{\delta}{\delta \bar{A}_{[a}^{A'B'}(x) }  e^{I(A,\bar{A},\xi, \bar{\xi})}) \ 
( \bar{\cal D}_a   \bar{\xi}^{B'}  \bar{\Phi} [ \bar{A}, \bar{\xi} ] )
( \frac{\delta}{\delta A_{b]}^{AB}(x) } {\cal D}_b \xi^B \Phi [A,\xi ] )
\nonumber 
\\
&=& -  \int dA d\bar{A} d\xi d\bar{\xi}   \int_\Sigma d^3 x \frac{1}{e} 
( n^{D}_{ B'} \frac{\delta}{\delta {A}_{[a A}^{D}(x) }  e^{I(A,\bar{A},\xi, \bar{\xi})}) \ 
( \bar{\cal D}_a   \bar{\xi}^{B'}  \bar{\Phi} [ \bar{A}, \bar{\xi} ] )
( \frac{\delta}{\delta A_{b]}^{AB}(x) } {\cal D}_b \xi^B \Phi [A,\xi ] )
\nonumber 
\\
&=& \int dA d\bar{A} d\xi d\bar{\xi}   \int_\Sigma d^3 x \frac{1}{e} 
( n^{D}_{ B'} e^{I(A,\bar{A},\xi, \bar{\xi})}) \ 
( \bar{\cal D}_a   \bar{\xi}^{B'}  \bar{\Phi} [ \bar{A}, \bar{\xi} ] )
(  \frac{\delta}{\delta {A}_{[a A}^{D}(x) }  \frac{\delta}{\delta A_{b]}^{AB}(x) } {\cal D}_b \xi^B \Phi [A,\xi ] )
\end{eqnarray}
We now integrate the $\bar{\cal D}_a$ by parts on $\Sigma$, which produces a boundary term 
\f
<R^{anti}>  =  <R^{anti}>^{boundary}+ <R^{anti}>^{bulk}
\ff
We deal with the bulk first
\begin{eqnarray}
<R^{anti}>^{bulk} &=&- \int dA d\bar{A} d\xi d\bar{\xi}   \int_\Sigma d^3 x 
( n^{D}_{ B'} e^{I(A,\bar{A},\xi, \bar{\xi})}) \ 
(    \bar{\xi}^{B'}  \bar{\Phi} [ \bar{A}, \bar{\xi} ] ) \frac{1}{e} 
(  \frac{\delta}{\delta {A}_{[a A}^{D}(x) }  \frac{\delta}{\delta A_{b]}^{AB}(x) } F_{ab}^{BE} \xi_E \Phi [A,\xi ] )
\nonumber
\\
&=&- \int dA d\bar{A} d\xi d\bar{\xi}   \int_\Sigma d^3 x
( n^{D}_{ B'} e^{I(A,\bar{A},\xi, \bar{\xi})}) \ 
(    \bar{\xi}^{B'}  \bar{\Phi} [ \bar{A}, \bar{\xi} ] )
 \frac{1}{e}  \hat{\cal C}_{DB} \xi^B \Phi [A,\xi ] ) \sim 0
\label{qconstraints}
\end{eqnarray}
where we use the second reality condition (\ref{R2}),
%\f
%< \Phi | {\cal D}_a (  \frac{\delta}{\delta {A}_{[a A}^{D}(x) }  \frac{\delta}{\delta A_{b]}^{AB}(x) } )  |\Phi > =0 
%\label{QT1}
%\ff

Eq. (\ref{qconstraints}) tells us that the quantum diffeomorpmism and Hamiltonian constraints are imposed with specific lapse and shift given be the Witten spinor, and only in the expectation value sense.
\f 
< \Phi | \int_\Sigma \bar{\xi}^{A'} n^{\ A}_{A'} \hat{\cal C}_{AB} \xi^B |\Phi > = 0 .  
\ff
In addition, note that we
find the constraints in a particular ordering
\f
\hat{\cal C}_{D}^{\ E}=  \frac{\delta}{\delta {A}_{[a A}^{D}(x) }  \frac{\delta}{\delta A_{b]}^{AB}(x) } F_{ab}^{BE}
\ff
%We impose the quantum reality condition, (\ref{R2}).  

Finally, we have
\f
- <R^{anti}>^{boundary} \equiv  <M_{ADM} > \geq 0
\ff
The operator is
\begin{eqnarray}
<M_{ADM} > &=&- \int dA d\bar{A} d\xi d\bar{\xi}   \int_{\partial \Sigma} d^2 \sigma_a   
( n^{D}_{ B'} e^{I(A,\bar{A},\xi, \bar{\xi})}) \ 
(    \bar{\xi}^{B'}  \bar{\Phi} [ \bar{A}, \bar{\xi} ] )  \frac{1}{e}
(  \frac{\delta}{\delta {A}_{[a A}^{D}(x) }  \frac{\delta}{\delta A_{b]}^{AB}(x) } {\cal D}_b \xi_B \Phi [A,\xi ] )
\nonumber
%\\
%&=&- \int dA d\bar{A} d\xi d\bar{\xi}   \int_{\partial \Sigma} d^2 \sigma_a  
%( n^{D}_{ B'} e^{I(A,\bar{A},\xi, \bar{\xi})}) \ 
%(      \bar{\Phi} [ \bar{A}, \bar{\xi} ] ) \frac{1}{e}
%(  \frac{\delta}{\delta {A}_{[a A}^{D}(x) }  \frac{\delta}{\delta A_{b]}^{AB}(x) }A_{b B}^D  \Phi [A,\xi ] 
%\nonumber
\\
& \geq &  0
\end{eqnarray}
where we use the boundary conditions (\ref{nullb0},\ref{s^an_a}).  This establishes the main result outlined in the introduction.

\section{Conclusions}

We conclude with some comments on future work.

\begin{itemize}

\item{}We so far have skirted the tricky issue of imposing asymptotically flat boundary conditions in the quantum theory.   This is possible because even the classical theory the proof works for a more general class of boundary conditions, establishing the positivity of the generalized $ADM$ energy (\ref{ADM1}).  

\item{}The above calculation establishes that a quantum positive energy theorem may be possible using a representation based on the Ashtekar connection.  Left open is a key question of whether this use of the Ashtekar connection is necessary or whether a quantum positive energy result can be achieved for representations based on other connections, i.e. for values of the Immirzi parameter besides $\gamma = \imath$.  One possible obstacle is that the Lorentzian Hamiltonian constraint is not polynomial for other values of $\gamma$, making the operator ordering and regularization issues much more challenging.   

\item{}Another important open question is whether there exist inner products which satisfy the reality conditions, (\ref{R1}, \ref{R2}) and positivity condition, (\ref{Q}).  

\item{}The form of the constraints needed for the result (\ref{<C>}) are very weak; it may be that  a stronger condition can be imposed.  However this cannot be that the ${\cal C}^{AB}$ annihilate the states, as those are not first class with the Witten equation (\ref{Wittenc2}).  Whether there is a stronger condition, consistent with  (\ref{Wittenc2}) is unknown.  

\item{}The Gauss law constraint does not come into the proof, except that the constraints found here are only equivalent to the $ADM$ Hamiltonian constraint and generators of spatial diffeomorphisms in the presence of the $SU(2)$ Gauss's law constraint, (\ref{Gauss}).  Thus we have to decide how Gauss's law is to be imposed in the quantum theory.  This is complicated by the fact that the Gauss law (\ref{Gauss}) does not commute under Poisson brackets with the Witten equation.  Thus we have three choices.  1) We can gauge fix and reduce, in which case the present results will have to be re-examined.
2)  We can impose the expectation value of the Gauss law constraint, following (\ref{<C>}), 
\f
<\hat{{\cal D}_a \tilde{E}} ^a_{AB} > =0
\ff
Or, 3),  we can extend the Gauss law  to act on the spinor $\xi^E$, to make it first class with the Witten equation,
$[{\cal G}_{AB}^{extend} , {\cal W}^E ] \approx 0$, where,
\f
{\cal G}_{AB}^{extend} = {\cal D}_a \tilde{E}^a_{AB} +\xi_{(A} \pi_{B)} 
\ff
and then impose it on a constraint on states
\f
{\cal G}^{AB}_{extended}  |\Phi > = 0
\ff
In this case we get a stronger constraint at a  cost of slightly weakening the equivalence of the Ashtekar constraints to the $ADM$ constraints.

\item{}This sketch of a formal proof should be strengthened by fully regulating the operator products involved. This can be attempted, either within the context of the kind of point split, but $SU(2)$ gauge invariant, regularization originally used in loop quantum gravity, as described in\cite{ls-92}, or the more rigorous approaches that have become standard since\cite{TT-book}.  This will, however, require that one key issue can be addressed:

\item{}{\it The issue of $\frac{1}{e}$}

Finally, we should comment on the problem of defining the inverse metric determinant operator $\frac{1}{e}$.
This is a crucial issue for loop quantum gravity and related non-perturbative approaches whose naive ground state
corresponds to $ < \tilde{E}_{AB}^a > \approx 0$.  The problem is that as, shown by \cite{Madhavan},  there exist
non-singular but degenerate solutions to the classical constraints of the Ashtekar formulation which are asymptotically flat but have negative $ADM$ energy.

We can note that in the classical proof, the antisymmetric part $\frac{1}{e}$ occurs in the combination
\f
 \frac{1}{e} \tilde{E}^{[a A}_D \tilde{E}^{b]}_{ AB} 
=
\epsilon^{abc}  e_c^{DB}
\ff
where $e_c^{DB}$ is the one form frame field.  In this case in loop quantum gravity we can use Thiemann's trick to write
\f
\hat{ e_c^{AB} (x)} = [  \hat{A}_c^{AB}( x) ,  \hat{V} ] 
\ff
where $\hat{V}$ is the volume operator and a regularization for the $ \hat{A}_c^{AB}( x)$ operator can be constructed from a limit of short holonomies, as explained in \cite{QSD1,TT-book}.

Using this the $ADM$ operator can be written in $LQG$ as 
\f
\hat{M}_{ADM} =  \int_{\partial \Sigma} d^2 \sigma_a  \epsilon^{abc}   [  \hat{A}_c^{AB}( x) ,  \hat{V} ] 
A_{b BA} 
\label{ADM-TT}
\ff
and the constraint operators, in the single densitized form,  are
\f
\hat{  \frac{1}{e} {\cal C}_{A}^{\ C}}=  \epsilon^{abc}   [  \hat{A}_{c A}^{B}( x) ,  \hat{V} ]  F_{ab B}^{\ \ \ C}
\label{C-TT}
\ff
To establish that this form of the constraints, (\ref{C-TT}),  lead to positivity of the corresponding form of the $ADM$ energy, 
(\ref{ADM-TT}), we must show that they are equivalent as operators to the forms that arise from squaring the Witten constraint.  That is, one must show the operator identity
\f
\hat{\frac{1}{e}}
\left (  \frac{\delta}{\delta {A}_{[a A}^{D}(x) }  \frac{\delta}{\delta A_{b]}^{AB}(x) } \right ) =  \epsilon^{abc} 
 [  \hat{A}_c^{DB}( x) ,  \hat{V} ] 
\ff
This is challenging.  Moreover,  I am not aware of a similar identity which can be used to define $\frac{1}{e} $ by itself or in
combination with the symmetric product of $\tilde{E}^a_{AB}$ which occur in the operator
$\hat{R}^{symm}$ in (\ref{<symm>}).   This remains the chief open problem required to run the proof in the  context
of loop quantum gravity.  

One promising approach is to modify loop quantum gravity to incorporate non-degenerate geometries along the lines of \cite{KS} or \cite{Bianca-rep}.

\end{itemize}

 \section*{Acknowledgements}

This work was begun in 2004 in collaboration with Artem Starodubtsev, and the main results in the classical proof and the basic strategy of the quantum proof (minus the idea of putting $\xi$ into the wave functional) were found with him, as
recorded in an unpublished draft \cite{ArtemLeepositive}.  

I am grateful to  Linqing Chen, Bianca Dittrich,  Ted Jacobson and to Madhavan Varadarajan and Miguel Campiglia for conversations and comments on drafts,  which were crucial for this work.
This research was supported in part by Perimeter Institute for Theoretical Physics. Research at Perimeter Institute is supported by the Government of Canada through Industry Canada and by the Province of Ontario through the Ministry of Research and Innovation. This research was also partly supported by grants from NSERC, FQXi and the John Templeton Foundation.

%======================================%
%<<<<<<<<<<<< BIBLIOGRAPHY >>>>>>>>>>>>%
%======================================%

\end{document}